# Title page

## Fully Automatic Content-Aware Tiling Pipeline for Pathology Whole Slide Images

Falah Jabar [1,2*], Lill-Tove Rasmussen Busund [2,3], Biagio Ricciuti [4], Masoud Tafavvoghi [5], Thomas K. Kilvaer [6,7], David J.Pinato [1,8], Mette Pøhl [9], Sigve Andersen [6,7], Tom Donnem [6,7], David J. Kwiatkowski [10,11], Mehrdad Rakaee [2,10,12]

[1] Dep. of Surgery and Cancer, Imperial College London, London, United Kingdom

[2] Dep. of Clinical Pathology, University Hospital of North Norway, Tromso, Norway

[3] Dep. of Medical Biology, UiT The Arctic University of Norway, Tromso, Norway

[4] Lowe Center for Thoracic Oncology, Dana-Farber Cancer Institute, Harvard Medical School, Boston, USA

[5] Dep. of Community Medicine, UiT The Arctic University of Norway, Tromso, Norway

[6] Dep. of Oncology, University Hospital of North Norway, Tromso, Norway

[7] Dep. of Clinical Medicine, UiT The Arctic University of Norway, Tromso, Norway

[8] Dep. of Translational Medicine, University of Piemonte Orientale, Novara, Italy

[9] Dep. of Oncology, Copenhagen University Hospital, Rigshospitalet, Copenhagen, Denmark

[10] Dep. of Medicine, Brigham and Women's Hospital, Harvard Medical School, Boston, US

[11] Dep. of Medical Oncology, Dana-Farber Cancer Institute, Harvard Medical School, Boston, US

[12] Dep. of Cancer Genetics, Oslo University Hospital, Oslo, Norway

*Corresponding author:

Falah Jabar, PhD

Imperial College London, Hammersmith Campus,

Du Cane Road, W12 0HS, London, UK.

E-mail addresses: f.rahim@imperial.ac.uk

**Abstract**—Tiling (or patching) histology Whole Slide Images (WSIs) is a required initial step in the development of deep learning (DL) models. Gigapixel-scale WSIs must be divided into smaller, manageable image tiles. Standard WSI tiling techniques often exclude diagnostically important tissue regions or include regions with artifacts such as folds, blurs, and pen-markings, which can significantly degrade DL model performance and analysis. This paper introduces WSI-SmartTiling, a fully automated, deep learning-based, content-aware WSI tiling pipeline designed to include maximal information content from WSI. A supervised DL model for artifact detection was developed using pixel-based semantic segmentation at high magnification (20x and 40x) to classify WSI regions as either artifacts or qualified tissue. The model was trained on a diverse dataset and validated using both internal and external datasets. Quantitative and qualitative evaluations demonstrated its superiority, outperforming state-of-the-art methods with accuracy, precision, recall, and F1 scores exceeding 95% across all artifact types, along with Dice scores above 94%. In addition, WSI-SmartTiling integrates a generative adversarial network model to reconstruct tissue regions obscured by pen-markings in various colors, ensuring relevant valuable areas are preserved. Lastly, while excluding artifacts, the pipeline efficiently tiles qualified tissue regions with minimum tissue loss.
In conclusion, this high-resolution preprocessing pipeline can significantly improve pathology WSI-based feature extraction and DL-based classification by minimizing tissue loss and providing high-quality – artifact-free – tissue tiles. The WSI-SmartTiling pipeline is publicly available on [GitHub](GitHub).

# 1. Introduction

Whole slide imaging enables us to the capture gigapixel whole slide images (WSIs) comprising the histological information previously only available on glass tissue slides. Subsequent processing of the WSI offers the potential of automation of tedious procedures like cell counting, diagnostic support for the pathologist, extraction of additional diagnostic/prognostic information and more. In recent years, deep learning (DL) models are used for detection, classification, and outcome prediction in various cancer types (Elemento et al., 2021; Van der Laak et al., 2021; Rakaee et al., 2024). To train DL models, the WSIs undergo preprocessing, normally include (Song et al., 2023; Yang et al., 2023): (1) Since WSIs are gigapixel images that cannot be processed or fed to the DL models at once, WSIs are divided into smaller, more manageable image tiles (**Fig. 1a**); (2) WSIs often contain pen-marking annotations (**Fig. 1b-c**), blurred areas (**Fig. 1d-e**), tissue folds (**Fig. 1f-g**) and non-tissue regions or backgrounds (**Fig. 1h-i**). To ensure reliable data for DL model training/testing, these artifacts and background regions must be identified and removed. Manually identifying and removing artifacts is impractical due to the need to visually evaluate gigapixels WSIs, making it extremely time-consuming for pathologists.

Including WSIs regions with artifacts and/or background in the development of any DL pipelines can significantly degrade the performance (e.g., overfitting, underfitting, poor generalization), potentially resulting in incorrect outcomes (Schömig-Markiefka et al., 2021;Wright et al., 2021;Weng et al., 2024). Therefore, tiling WSIs is an essential preprocessing step that must fulfill three key requirements:

1) High magnification – WSIs should be tiled at high magnification to ensure high-quality data for DL models. This is a prerequisite for many pathology foundation models used in feature extraction (Neidlinger et al., 2024).
2) Artifact-free tiling – Only qualified tissue regions should be tiled, while artifacts and background areas should be excluded. This not only enhances data quality but also optimizes computational efficiency by eliminating uninformative regions.
3) Maximized tissue inclusion – The tiling process should minimize the exclusion of diagnostically relevant tissue regions to preserve as much useful information as possible.

Although tiling is typically performed as shown in **Fig. 1a**, conventional approaches often result in the exclusion of important tissue regions. Exclusion typically occurs for one of three reasons: (1) regions near tissue boundaries, e.g. tiles 6, 7, 10, and 23 in **Fig. 1a**, are excluded due to a high percentage of background. However, they may contain portions of good quality tissue important for diagnosis; (2) regions containing artifacts, e.g. tiles 12 and 17 in **Fig. 1a**, such as tissue folds are excluded, but may contain a high percentage of good quality tissue, (3) regions with pen-marking, e.g. tile 25 in **Fig. 1a**, may be excluded but contain good amounts of quality tissue (see also **Fig. 1c**). Loss of good quality images due to exclusion can be minimized by using more efficient tiling techniques, including adjusting tile positions to avoid background and artifacts while including the good quality regions.

The primary objective of this paper is to develop a preprocessing pipeline that curates and prepares WSIs for training DL models, particularly for quantitative tasks such as cell detection and classification. Artifacts present in WSIs can lead to over- or under-detection of cells and negatively impact cell classification accuracy (Weng et al., 2024). To address this, the proposed pipeline integrates multiple preprocessing steps to retain high-quality image data while eliminating artifacts. The main achievements are as follows:

1) Development of a large and diverse dataset of WSIs artifacts, including tissue folds, blurring, and background, obtained from multiple WSIs and annotated by experts.
2) Development of a fully automated DL model that uses pixel-based segmentation with high resolution to classify WSIs regions into categories such as qualified tissue, folding, blurring or background.
3) Integration of Generative Adversarial Networks (GANs) model to reconstruct tissue regions obscured by pen markings, removing marker signs where possible, while preserving relevant tissue tiles (Ramanathan et al., 2023).
4) Implementation of an efficient, content-aware tiling procedure that maximizes the number of qualified tiles, achieving retention of maximal tissue features.

The rest of this paper is organized as follows: Section 2 reviews related work; Section 3 describes the methods and proposed WSI-SmartTiling pipeline; Section 4 presents the performance evaluation results; Section 5 concludes the paper with future directions.

# 2. Related Work

Early work on artifact detection relied on image feature extraction and color processing. In (Gao et al., 2010;Wu et al., 2015), a combination of image contrast, gradients, and pixel statistics, was used to detect blurry regions in WSIs. Similarly, in (Hashimoto et al., 2012), image noise information and sharpness were combined to detect blurring artifact in the WSIs. In (Palokangas et al., 2007; Bautista et al., 2009; Kothari et al, 2013; Mercan et al., 2016), image color and texture information were used to identify folding artifacts in WSI. Although these approaches laid a foundation for WSI artifact detection, they were limited in handling WSIs with different characteristics including variations in colors, staining methods, tissue and artifact types. In (Albuquerque et al., 2021), multiple DL pipelines were trained and compared for detecting blurred regions in WSIs.

In (Kohlberger et al., 2019), ConvFocus was developed to quantify the severity and localize blurry regions in WSI. Similarly, in (Babaie et al., 2019), multiple DL pipelines and binary classifiers were trained and compared for detecting folded regions in WSIs. In (Kanwal et al., 2023), a vision transformer is trained to detect air bubbles for diagnosis purposes. In (Kanwal et al., 2022), a DL pipeline is proposed to assess the impact of color normalization over blood and fold tissue detection. All these publications relied on training a single pipeline to detect one or two artifacts. In addition, (Kanwal et al., 2022) and (Kanwal et al., 2023) were trained and tested on a limited single-cohort dataset consisting of 55 WSIs, without validation on external datasets.

In (Ali et al., 2019), a convolutional Neural Network (CNN) based classifier was employed to detect WSIs regions with pen-markers, followed by a generative model to remove pen-markers. In (Jiang et al., 2020; Venkatesh et al., 2020; Ramanathan et al., 2023), GAN models were used to restore WSI regions covered by pen-markers.

In (Kanwal, et al., 2024a; Kanwal, et al., 2024b), DL pipelines were proposed to predict the presence of artifacts in tiled WSIs, and then classify tiles into categories such as damaged tissue, blur, folded tissue, air bubbles, and blood. Both approaches were trained on a limited single-cohort datasets. In addition, they did not detect regions with pen-markings. In a most recent report, (Weng et al., 2024), GrandQC was proposed for artifact detection in WSIs. However, it was trained only on WSIs with 5x, 7x, and 10x magnification, limiting its performance on higher magnifications like 20x or higher. Both PathProfiler (Haghighat et al., 2021) and HistoROI (Patil et al., 2023) had similar limitations.

Several preprocessing toolboxes such as TiaToolbox in (Pocock et al., 2022), PathML in (Berman et al., 2021), and CLAM in (Lu et al., 2021), have been developed for computational pathology. These toolboxes provide functionalities such as WSI tiling, tissue-background separation, and application programming interfaces (APIs) to support the development and integration of DL models into computational pathology workflows. However, they were not designed for artifact detection. Among these only PathML had the ability to detect pen-markings, but not other types of artifacts.

In summary, these existing methods do not provide a complete, content-aware preprocessing pipeline capable of tiling WSIs at high magnification while effectively excluding artifacts and minimizing the exclusion of qualified tissue. In addition, most rely on tile-level classification, instead of pixel level, potentially discarding quality images from some tumor regions. Furthermore, many approaches are trained on limited datasets focused on low magnification (e.g., 10x), limiting their utility for high quality (20x or higher) WSI magnification.

## 3. Method

**Fig. 2** illustrates the architecture of the proposed WSI-SmartTiling. Briefly, the workflow starts by identifying the region of interest (WSI-ROI) and dividing it into smaller image tiles. Pen-marking detection is then applied to categorize the tiles into two classes: those with high pen-marking (which are discarded) and those with medium and low pen-marking. Tiles with medium and low pen-marking undergo a pen-marking removal process, resulting in clean image tiles. Next, the clean image tiles are fed into the proposed artifact detection model to identify artifacts, followed by an optimization technique to select the best tiles—those with minimal artifacts and background and maximum qualified tissue. Finally, WSI are reconstructed by combining the selected tiles to generate the final output. Additionally, the model generates a segmentation for the entire WSI and provides statistics on tile segmentations. Pipeline development steps and technical details are provided in the following sections.

### 3.1. Data Materials and Annotation

#### 3.1.1. Development Dataset

Hematoxylin and Eosin (H&E) slides were collected from multiple centers with different staining procedures. The dataset includes: (1) 446 WSIs of metastatic non-small cell lung cancer (NSCLC) patients from the Dana-Farber Cancer Institute (DFCI), of which 66 had artifacts (Rakaee et al., 2023a) and (2) 453 WSIs of NSCLC patients from a Scandinavian multi-institutional clinical trial (TNM-I), of which 73 had artifacts (Rakaee et al., 2023b). The WSIs from DFCI were scanned using the Aperio ScanScope AT (Leica Biosystems GmHB, Nussloch Germany) scanner at 0.5 μm/px (20x magnification), while the WSIs from TNM-I were scanned using the 3DHistech Pannoramic Flash III (3DHistech Ltd. Budapest Hungary) scanner at 0.25 μm/px (40x magnification). In total, 24,942 tiles were generated, with the following distribution: 9,828 artifact-free (qualified tissue) tiles, 7,334 tiles with folding, and 7,780 tiles with blurring (**Fig.S1-a** in the supplementary materials). Additionally, 1000 background tiles from various WSIs were included to ensure the model can detect backgrounds with varying colors. The dataset was built using the procedure described in section 3.1.3. All the patients were consented and the data collection was ethically approved by the Institutional Review Board at each participating institute: DFCI-2021P000557; TNM-I-REK2016/2054.

#### 3.1.2. External Validation Dataset

This external dataset includes 18 H&E-stained WSIs, which had artifacts, including six different organs: brain, breast, bladder, lung, liver and kidney, obtained from the TCGA dataset (National Cancer Institute, 2024). Each organ is represented by three WSIs (3 WSIs per organ, totaling 18). The brain and bladder WSIs were scanned at 0.5 μm/px (20x magnification), while the remaining organs were scanned at 0.25 µm/px (40x magnification). In total, 10,701 tiles were generated, with the following distribution: 3117 artifact-free (qualified tissue) tiles, 1350 tiles with folding, and 6243 tiles with blurring (**Fig.S1-b** in the supplementary materials).

*3.1.3. Annotation Procedure*

Supervised pixel-based segmentation models require ground truth data where each pixel is accurately labeled, making the WSI annotation task challenging and time-consuming. Consider the WSI-ROI (as shown in **Fig. 2**) as a 2D grid of pixels, $I$, where each pixel is indexed by $(i,j)$ and represented by three color values (R, G, B). The set of labels is $L = \{l_k\}$, where $k = 0,1,2,...,M$, and $M$ is the number of distinct labels. The objective is to assign a label $l_k$ to each pixel within in a region indexed by $r$, where $r = 1,2,3,...,R$, and $R$ is the number of distinct regions. For example, label 0 is assigned to all pixels in the background regions. Thus, $I(i,j)_r = l_k$, where $l_k \in L$. To achieve this, an annotation procedure was proposed to label WSI pixels specifically for segmentation tasks. A schematic overview of the annotation process is provided in **Fig. S2** of the supplementary materials, and is summarized as follows:

- **Annotation**: Set magnification in QuPath software (Bankhead, et al., 2017) to a minimum of 20x. Manually annotate regions with artifacts (e.g., folds or blurring) within the QuPath environment to ensure precise labeling. Next, tile the annotated regions of WSI to generate tiles from regions containing artifacts. Each tile has a resolution of $(H_t, W_t, 3)$ pixels, where 3 represents the (R, G, B) channels. The generated tiles include both unannotated regions (qualified tissue and background) and annotated regions (folding and blurring). For each tile, a corresponding 2D segmentation mask is generated with the same $(H_t, W_t)$. In the mask, unannotated regions are labeled as 0, folding is labeled as 2, and blurring is labeled as 3.
- **Post-processing**: Using the information in the segmentation mask, the unannotated regions in each RGB tile are separated from the annotated regions. The unannotated regions are further processed to distinguish background from qualified tissue by applying the background segmentation technique proposed in (Abrol et al., 2023). In short, the RGB image tile is converted to hue, saturation, and lightness (HSL) color space. Then, color thresholding is applied to separate the background from the tissue regions. The final segmentation mask is then generated, where the background is labeled as 0, qualified tissue as 1, folding as 2, and blurring as 3 (**Fig.S1-c** of the supplementary materials).

*3.2. Model development*

*3.2.1. WSI Tiling*

WSIs typically contain large backgrounds outside tissue regions, and tiling these regions is computationally inefficient and time-consuming. To exclude background, the WSI with resolution $(H_{WSI}, W_{WSI})$ pixels is first down sampled to (5000, 5000) pixels, and then Otsu segmentation (Open Source Computer Vision, 2024) is applied to generate a tissue mask, separating tissue regions (labeled as 255) from the background (labeled as 0), as shown in **Fig. 2**. Small tissue regions that have less than 25 pixels are removed, followed by a dilation morphological dilation (Scikit Image, 2024), using 8-neighbor connectivity to expand the tissue borders. The identified tissue regions (WSI-ROI), indicated by the purple rectangle in **Fig. 2**, have a resolution of $(H_{ROI}, W_{ROI})$ pixels. The tiling is then performed over WSI-ROI. Tiling begins from the top-left corner of WSI-ROI, using a tile size of $(H_t, W_t)$ pixels, with no overlap. Each tile is denoted as $t_n^c$, where $n = 1, 2, 3, ..., N$, with $N$ being the total number of non-overlapped tiles. For each tile, four neighbor tiles are extracted, left-tile ($t^l$), up-tile ($t^u$), right-tile ($t^r$), and down-tile ($t^d$) as shown in **Fig. 2**. These neighboring tiles overlap the center tile ($t^c$), with the degree of overlap controlled by a parameter denoted $O_t$, for example, $O_t = 25\%$. As a result, several sets of image tiles are generated, indexed as $s_1, s_2, ..., S_N$. Each set contains five tiles: the center tile and its four neighboring tiles, yielding a total of 5x$N$ tiles. To speed up the process, tiling is performed using parallel processing (Pytorch, 2024a).

*3.2.2. Pen-marking Detection and Removal*

An automatic procedure was implemented to exclude tiles with high pen-marking. For each set of tiles and for each tile, the percentages of red, green, blue, and dark pixels were calculated separately, to identify the presence of specific pen-markers. These percentages are denoted as $P_{red}, P_{green}, P_{blue}, P_{dark}$, representing the amount of red, green, blue, and dark pixels respectively within the range [0,1], where valued close to 1 means high amount of pen-marking, and values close to 0 means low amount of pen marking. Based on these percentages, tiles were classified into three categories:

- High pen-marking: A tile is classified as having high pen-marking if $P_{red}, P_{green}, P_{blue}, P_{dark}$ is greater than a predefined threshold $P_{max}$ (e.g., $P_{max} = 0.9$).
- Medium pen-marking: A tile is classified as having medium pen-marking if $P_{red}, P_{green}, P_{blue}, P_{dark}$ is less than a predefined threshold $P_{max}$ and greater than $P_{min}$ (e.g., $P_{min} = 0.2$).
- Low pen-marking: A tile is classified as having low pen-marking if $P_{red}, P_{green}, P_{blue}, P_{dark}$ is less than a predefined threshold $P_{min}$.

Tiles with high pen-marking are excluded, while those with medium or low pen-marking are cleaned, using pen-marking removal pipeline proposed in (Ramanathan et al., 2023). Briefly, this pipeline consists of a two-stage network: a ResNet-18 (He et al., 2016) based binary classifier for pen-marking detection and filtering, followed by a Pix2Pix network for pen-marking removal (Isola et al., 2017). The model has been trained and tested on a diverse dataset containing multiple pen-marker colors, demonstrating strong performance on independent datasets. After the detection and removal of pen-marking, the tile sets $s_1, s_2, \ldots, S_N$ may contain less than five tiles. In some cases, entire tile sets are discarded if the central tile and its four neighboring tiles are all classified as having high pen-marking.

*3.2.3. Artifact Segmentation Model*

After removing pen-marking from the tiles, artifacts such as blurring, folds, or background regions need to be detected if present. To address this, a pixel-based semantic segmentation model was proposed. The artifact detection model is a dual-branch hierarchical global–local fusion network built based on the work in (Wang et al., 2023). Briefly, it combines a global encoder, Swin Transformer – a vision transformer that extracts global image features – and a local encoder, ConvNeXt – a convolutional neural network that extracts local image features. This model is designed to handle segmentation tasks in WSIs. The Swin Transformer focuses on global image structure, while ConvNeXt refines local details, ensuring high segmentation accuracy. The model architecture is provided in **Fig.S3** of the supplementary materials. The model is trained on the development dataset described in section 3.1.1, with the training procedure explained in the following section.

The input tiles, each with a size of 270 by 270 pixels, are fed into the model, which generates pixel-based segmentation for each tile, assigning labels to each pixel: 0 for background (black color), 1 for qualified tissue (green color), 2 for fold (red color), and 3 for blurring (orange color). In the post-processing step, small background regions in the tile segmentation mask (holes with fewer than 25 pixels) are closed.

*3.2.4. Model Training*

The development dataset was randomly split, 87% for training and 13% for testing. Detailed information regarding the dataset split can be found in **Fig.S1-a** in the supplementary materials. The Adam optimizer was employed to minimize the loss function and achieve optimal performance. Various batch sizes and learning rates were tested during the optimization process. Specifically, learning rates were evaluated in the range of $10^{-2}$ to $10^{-4}$, and batch sizes of 16, 32, and 64 were tested. The optimal configuration that better minimizes the loss function was found to be a batch size of 64 and a learning rate of $10^{-3}$.

Throughout the training process, Weights and Biases (WandB), a widely used tool for tracking, visualizing, and optimizing machine learning experiments (Weights & Biases, 2024), was employed to monitor the model's progress. To enhance the model's generalization, various data augmentation was applied, including horizontal and vertical flips and color augmentation (PyTorch, 2024b). Horizontal and vertical flips were used to increase spatial diversity, while color augmentations (brightness = 0.5, contrast = 0.5, saturation = 0.5, hue = 0.5) were applied to improve robustness against color variations. Initial parameter settings for the network were based on guidance from (Wang et al., 2023). The model was trained for 150 epochs, as increasing the number of epochs beyond this did not lead to further performance improvements. The training was performed on a system equipped with an NVIDIA A100 GPU (80GB of memory) running CUDA version 12.2. The system also featured an Ubuntu operating system (version 20.4), a 40-core CPU, and 512GB of RAM.

*3.2.5. Tile Selection*

Once the artifact segmentation is completed, for each tile set $s_1, s_2, \ldots, S_N$, the best qualified tile is selected by applying the following steps:

• Compute artifact percentages: for each tile $t^c, t^l, t^u, t^r, t^d$ in the set, the percentage of artifacts is computed and normalized to the range [0,1]. These percentages are denoted as $P_{fo}, P_{bl}, P_{bg}$, representing the percentage of fold, blur, and background, respectively.

• Compute cost function: For each $t^c, t^l, t^u, t^r, t^d$ in the set, compute the cost function by applying:

$$\boldsymbol{C} = \lambda_{fo}P_{fo} + \lambda_{bl}P_{bl} + \lambda_{bg}P_{bg}, \tag{1}$$

where $\boldsymbol{C}$ is a vector conaing the cost function values for $t^c, t^l, t^u, t^r, t^d$; $\lambda_{fo}, \lambda_{bl}, \lambda_{bg}$ are the weighting parameters for folds, blurring, and background respectively. For this study, all weighting parameters are set to 1.

• Tile selection: select the tile with the minimum amount of artifacts and background using:

$$t^{sel} = \min(\boldsymbol{C}). \tag{2}$$

Applying the above steps optimizes the tiling of WSI by: (1) ensuring only qualified tissue regions are tiled; (2) avoiding regions with significant artifacts, (3) minimizing the exclusion of qualified tissue (**Fig. 2**).

*3.3. WSI-SmartTiling Model Evaluation*

A subjective evaluation was conducted to compare the model's performance with expert pathologists. Ten WSIs (WSI-1 to WSI-10) were selected from the development test set. For each tiled WSI, 10 sets of tiles were randomly selected, each consisting of five overlapping tiles (25% overlap): center, left, up, right, and down. In total, 100 sets were generated (10 WSIs × 10 sets per WSI). A graphical user interface (GUI) was used to display the image tiles, allowing expert pathologists to select the best-qualified tile (i.e., minimal artifacts, minimal background, and maximum qualified tissue). If no tile met the criteria, pathologists could select "None."

Two investigators (L.T.B, S.A) independently evaluated the tiles. The test instructor reviewed their selections to resolve disagreements (no disagreements were detected), and a final selection was recorded for each set. The model's selection was then compared to the expert selection to assess its accuracy. To evaluate the ability of WSI-SmartTiling to preserve qualified tissue, its performance was compared to the standard tiling technique (**Fig. 1a**). The percentage of qualified tissue (QT) gain was computed using the following formula:

$$\frac{QT\% \text{ for proposed tiling} - QT\% \text{ for standard tiling}}{QT\% \text{ for standard tiling}}. \quad (3)$$

## 4. Results: Performance Evaluation and Benchmark

### 4.1. Evaluation of Artifact Detection Model

The performance of artifact detection model was evaluated using the test set from the development dataset as well as the external validation set. The state-of-the-art methods used for comparison include five recently proposed pipelines: GrandQC (Weng et al., 2024) pixel-wise segmentation model developed for artifact detection, and four tile-wise classification model with different network architectures (Kanwal, et al., 2024b), namely MoE-CNN, MoE-ViT, multiclass-CNN, and multiclass-ViT. Quantitative metrics, including total accuracy (Acc), precision, recall, and F1 score, were used to evaluate classification performance, and Dice metric used to evaluate segmentation performance (Wang et al., 2023).

### 4.2. Quantitative Evaluation

**Fig. 3** shows the classification performance of the proposed and benchmark pipelines (Kanwal et al., 2024b) on the development and external validation datasets. Confusion matrices are provided in **Table S.1**. The proposed model outperformed all benchmarks across artifact types, achieving precision, recall, F1-score, and total accuracy above 95%, demonstrating strong generalizability. MoE-CNN, previously the best-performing model (Kanwal et al., 2024b), showed lower accuracy (74% development, 70% external) and struggled with the fold class. Its precision, recall, and F1-scores averaged 74%, 70%, and 71% in development and 68%, 72%, and 65% in external validation. MoE-ViT showed inconsistent performance, with 52% accuracy in development but 76% in external validation. It underperformed in artifact-free and fold classes in development but improved externally, suggesting poor generalization. Multiclass-CNN performed similarly to MoE-CNN in development (75%) but improved in external validation (83%). While its precision, recall, and F1-score were 82%, 70%, and 69% in development and 82%, 73%, and 74% externally, it also struggled with the fold class. Multiclass-ViT achieved 58% accuracy in development and 67% in external validation, with weak performance in the blur class. Its metrics improved externally but remained inconsistent across artifact types, similar to MoE-ViT.

To rule out organ-specific effects on model performance, an external dataset comprising six different organs – bladder, brain, breast, kidney, liver, and lung – was used for validation. The proposed model was benchmarked against the best-performing state-of-the-art model, MoE-CNN, and outperformed it across all metrics (**Fig. S4**, **Table S2**).

GrandQC is the most recent model proposed for artifact detection (Weng et al., 2024). To assess comparative performance, the proposed model achieved an average Dice score of 96% on the development dataset and 94% on the external validation dataset, demonstrating strong generalization. In contrast, GrandQC achieved an average Dice score below 50% on both datasets, likely due to its training on WSIs at low magnifications (5x, 7x, 10x), which significantly limits performance at higher magnifications (**Fig. 4**).

### 4.3. Qualitative Evaluation

To visually compare tile segmentation and classification performance, the proposed model was evaluated against MoE-CNN and GrandQC across six organs (bladder, brain, breast, kidney, liver, lung) and background-only regions (**Fig.5**). The proposed model achieved high accuracy in artifact detection and tile classification, with predicted labels matching the ground truth across all organ and background tiles. It effectively distinguished qualified (artifact-free) from unqualified (artifact-containing) tiles.

MoE-CNN, operating at the tile level rather than individual pixels, showed limited artifact detection accuracy. It frequently misclassified qualified tiles as containing folds, particularly in brain, kidney, liver, and lung samples (**Fig. 5**). This misclassification likely results from color similarities between certain cell regions and fold artifacts. Additionally, some background-only tiles, particularly reddish ones, were incorrectly classified as qualified.

GrandQC exhibited poor segmentation performance across all organs, struggling with high-magnification WSIs. It often misidentified qualified tissue as blurred regions and over-detected folds. However, it performed better than MoE-CNN on background-only tiles (**Fig. 5**). Additional benchmark segmentations on representative WSIs from different tissue types and magnifications (20x and 40x) are provided in the supplementary materials (**Fig. S5**, **Fig. S6**).

*4.4. Evaluation of WSI-SmartTiling Pipeline*

A subjective test was conducted to compare the model's tile selection with that of expert pathologists (**Method 3.3**). The model achieved at least 80% accuracy, correctly selecting tiles in at least 8 out of 10 sets (**Fig. 6a**). Misclassifications occurred when multiple tiles contained similarly qualified tissue, making distinctions difficult even for pathologists.

To assess tissue preservation compared to conventional tiling models, we calculated the percentage of qualified tissue gain per WSI (**Method 3.3**). Gains ranged from 0.76% to 7.5% (**Fig. 6b**), indicating improved tissue retention, particularly in artifact-prone regions and near tissue borders. This effect was most pronounced in fragmented tissue regions (**Fig. 7**, **Fig. S7**).

*4.5. Comparison with other Pipelines*

**Table 1** summarize and compare the features offered by commonly used WSIs processing pipelines – GrandQC (Weng et al., 2024), MoE-CNN (Kanwal, et al., 2024b), PathProfiler (Haghighat et al., 2021), Pathml (Berman et al., 2021), HistoROI (Patil et al., 2023), CLAM (Lu et al., 2021), TiaToolBox (Pocock et al., 2022) – and the proposed WSI-SmartTiling. The comparison focuses on support for tissue masking, folding, blurring, pen-marking detection, pen-marking removal, Content-Aware (CA) tile classification, tiling optimization, maximum magnification (MM) support, and targeted application. While most of these pipelines are designed for computational pathology or WSI analysis, WSI-SmartTiling is specifically developed for efficient preprocessing of WSIs. It is worth noting that different applications have different requirements. For example, preprocessing pipelines aimed at cleaning WSIs must support high magnifications to ensure accuracy and effectiveness.

We compared WSI-SmartTiling to various pipelines based on their artifact handling capabilities (**Fig. 8**). Among benchmarked models, GrandQC and MoE-CNN are specifically designed for artifact detection in WSIs; however, as previously observed at the ROI level (**Fig. 5**), both exhibit poor accuracy in identifying folding artifacts at the WSI level. PathProfiler and CLAM generate scoring/attention maps to highlight important WSI regions but lack artifact detection capabilities. PathML produces a segmentation mask that separates tissue from the background and detects pen-marked regions, but it cannot identify other artifacts. Additional WSI examples are provided in **Fig.S8** of the supplementary materials.

## 5. Discussion and Conclusion

This paper presents a robust artifact detection model, and WSI-SmartTiling, a fully automated, content-aware tiling pipeline designed to perform robust histopathology image preprocessing. First, an artifact detection segmentation model was developed to address key challenges in histopathology, particularly the identification and exclusion of WSI regions containing artifacts such as tissue folds, blurring, and background. The model employs a dual-branch hierarchical global–local fusion network, combining convolutional neural networks (CNNs) and a vision transformer, to achieve superior artifact detection performance. Quantitative and qualitative evaluations on diverse internal and external datasets—including WSIs from various tissue types—demonstrated that the model outperforms state-of-the-art methods in accurately identifying and segmenting artifact regions.

Second, WSI-SmartTiling was introduced as a fully automated, content-aware pipeline for optimizing WSI tiling. The pipeline integrates efficient tiling techniques, pen-marking detection and removal, artifact detection, and an optimization procedure to ensure that only qualified tissue regions are selected while excluding artifacts. This optimized tiling approach is particularly crucial in histopathology, where many routine WSIs are from biopsy samples with limited tissue content. By maximizing tissue retention while minimizing artifacts, WSI-SmartTiling provides high-quality input for downstream deep learning applications.

Our study has some limitations. First, the training dataset primarily consisted of lung cancer WSIs, including cases with metastatic sites/organs, and has been validated on external datasets from different tissue types. However, some cancer types may contain structures that resemble artifacts. For example, melanophages in melanoma, due to their brown-to-black melanin granules, could be misclassified as marker signs. Future work should include testing the model on such cases to improve generalizability. Second, the model was trained using two widely used image formats (SVS and MRXS) at different magnifications. However, other scanners/formats such as NDPI, VSI, etc have not been explored and should be considered in future evaluations to ensure broader applicability.

Overall, our pipeline can be used as a standalone tool for generating high-quality WSI tiles or integrated into upstream stages of computational pathology workflows such as CLAM (Lu et al., 2021), STAMP (El Nahhas et al., 2025), etc, for efficient preprocessing. By improving artifact removal and optimizing tiling, WSI-SmartTiling can significantly enhance the quality of histopathological image analysis in deep learning applications.

**Declaration of competing interest**

Dr Ricciuti reported personal fees from Amgen, AstraZeneca, Regeneron, and Bayer outside the submitted work. Dr Pinato reported consulting fees from ViiV Healthcare, Bayer, Roche, MiNa Therapeutics, Mursla Bio, H3B, AstraZeneca, Da Volterra, Exact Sciences, Ipsen, Avammune, Lift Biosciences, Starpharma, Boston Scientific, Eisai, Bristol Myers Squibb, Merck Sharp & Dohme, and GlaxoSmithKline outside the submitted work. Dr Kwiatkowski reported support from Genentech, AADI, and Revolution Medicines; and consulting fees from Genentech, AADI, Expertconnect, Guidepoint, Bridgebio, Slingshot Insights, William Blair, MEDACorp, and Radyus Research, all outside the submitted work. Dr Rakaee reported honoraria from AstraZeneca. No other COIs were reported.

**Funding sources**

This research is supported by the Northern Norway Regional Health Authority (grant HNF1660-23) and the Norwegian Cancer Society (grant 273190-2024). The funders had no role in the design and conduct of the study; and decision to submit the manuscript for publication.

**Data/code availability**

The annotated data/tiles will be available for research purposes upon request. The TCGA datasets used in this study are openly and publicly available at https://portal.gdc.cancer.gov/. The code is available on GitHub.

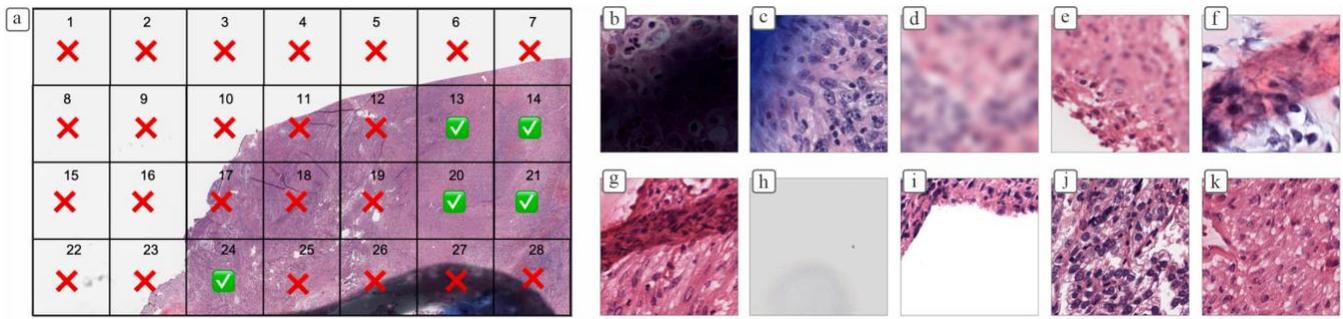

**Fig. 1. Tiling and tissue artifacts**: a) WSIs standard tiling: Tiles marked with ❌ indicate unqualified tiles due to artifacts and/or too much background, while tiles marked with ✅ indicate qualified tiles; b)-k) are examples of image tiles obtained from different WSIs: b)-c) image tiles with black and blue pen-marking; d)-e) image tiles with blurring artifact; f)-g) image tiles with folding artifact; h) image tile with only background and i) image tile with tissue and too much background; j)-k) qualified (artifact-free) image tiles.

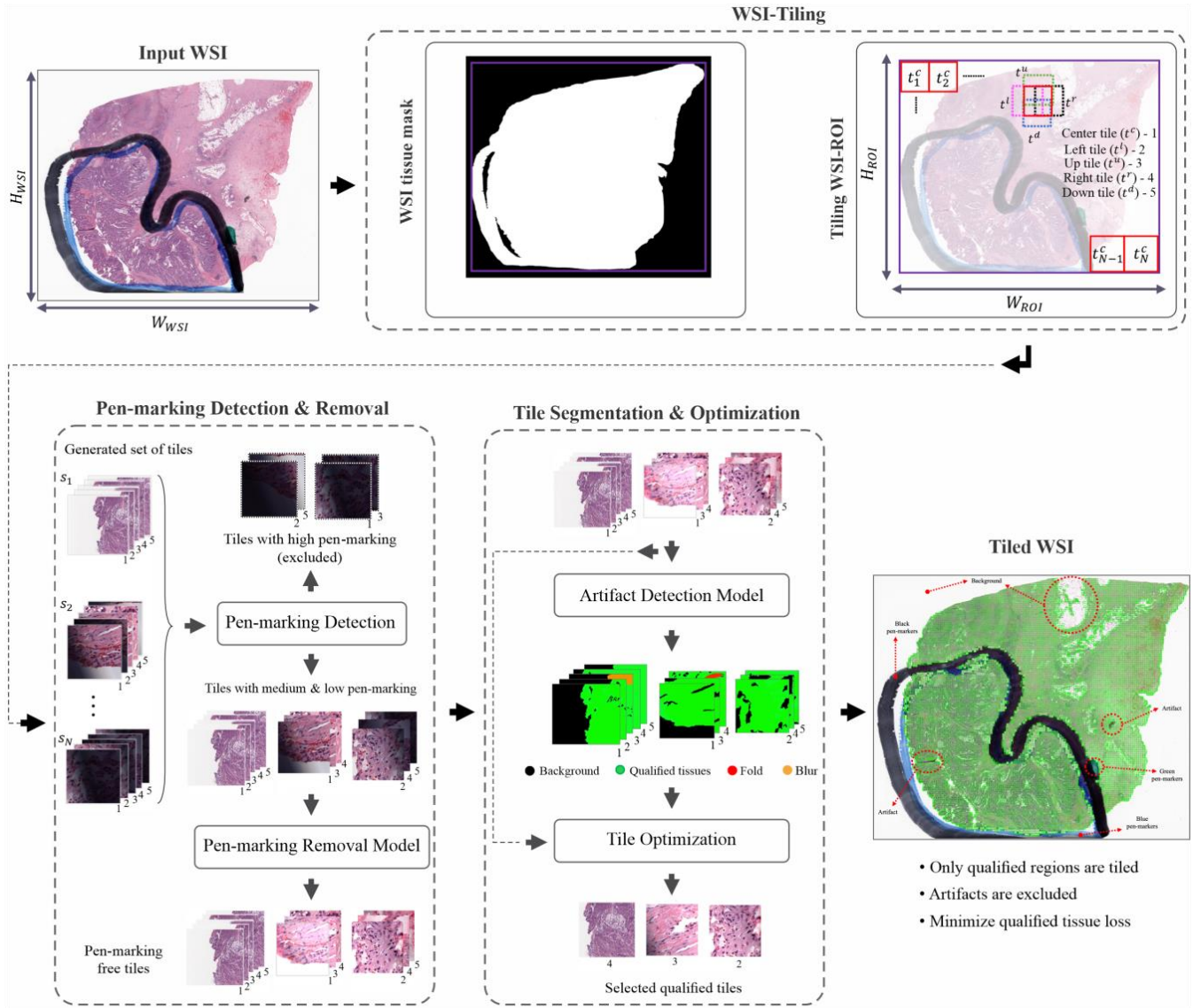

**Fig. 2. Process workflow**: Architecture of the proposed WSI-SmartTiling pipeline.

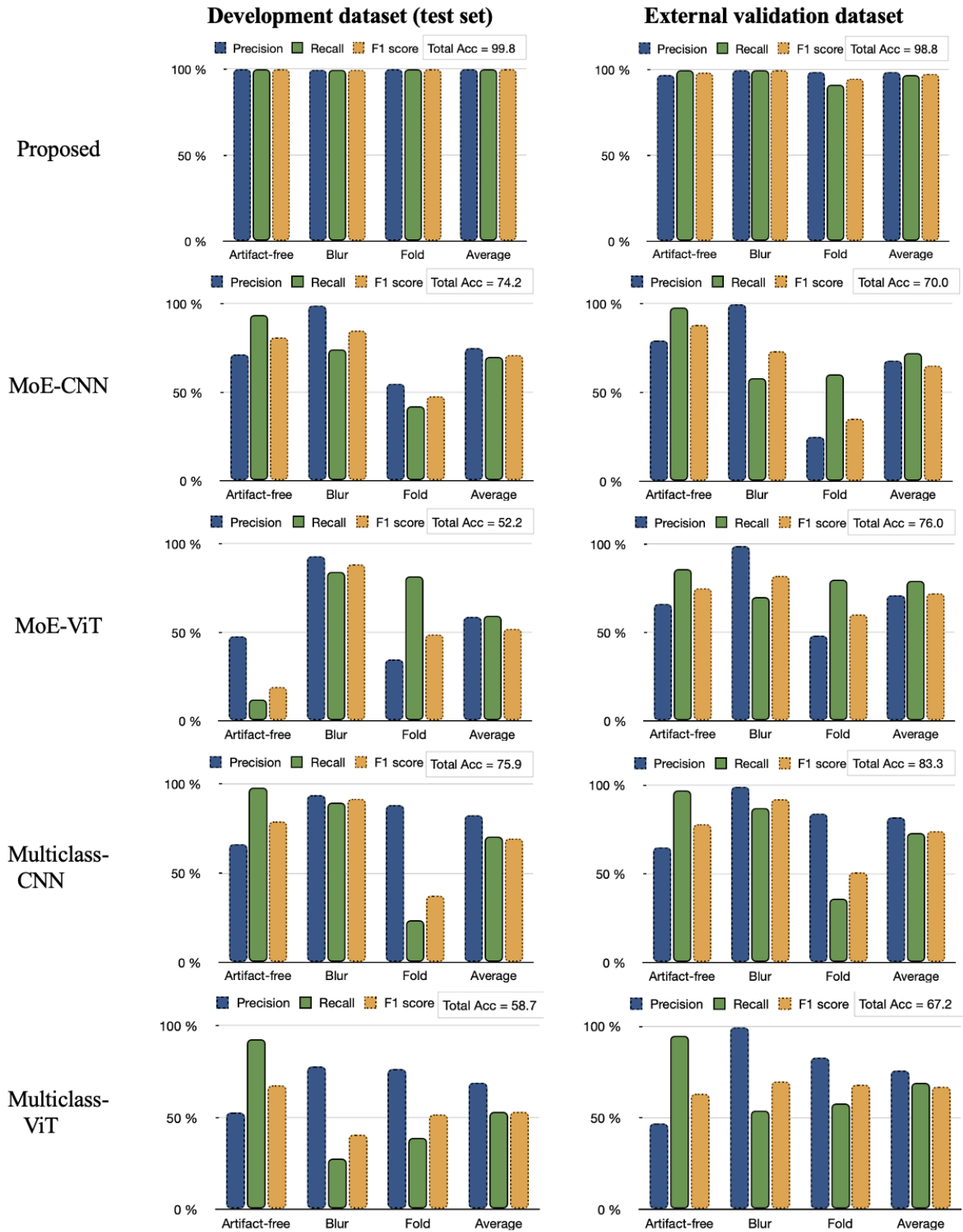

**Fig. 3. Classifier performance:** Classification metrics for the proposed artifact detection model compared with the state-of-the-art methods (Kanwal, et al., 2024b). The left column shows performance metrics on the development dataset (test set), while the right column presents performance metrics on the external validation dataset. The last three columns represent the average performance across artifact-free, blur, and fold categories.

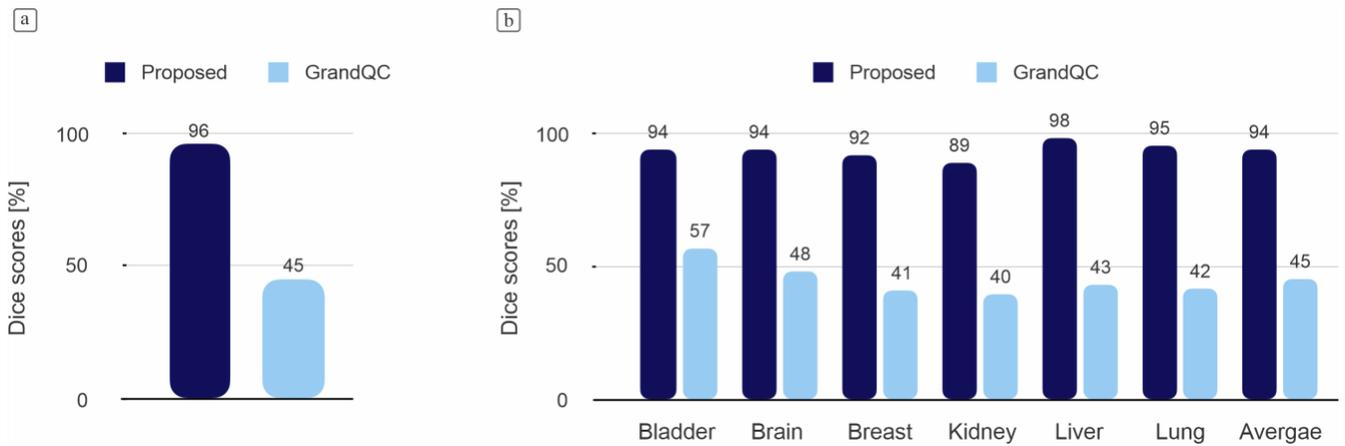

**Fig. 4. WSI-SmartTiling vs. GrandQC performance**: Segmentation metrics for the proposed artifact detection model compared with GrandQC (Weng et al., 2024): a) Development dataset (test set); b) External validation dataset (the last two columns represent the Dice scores averaged across organs).

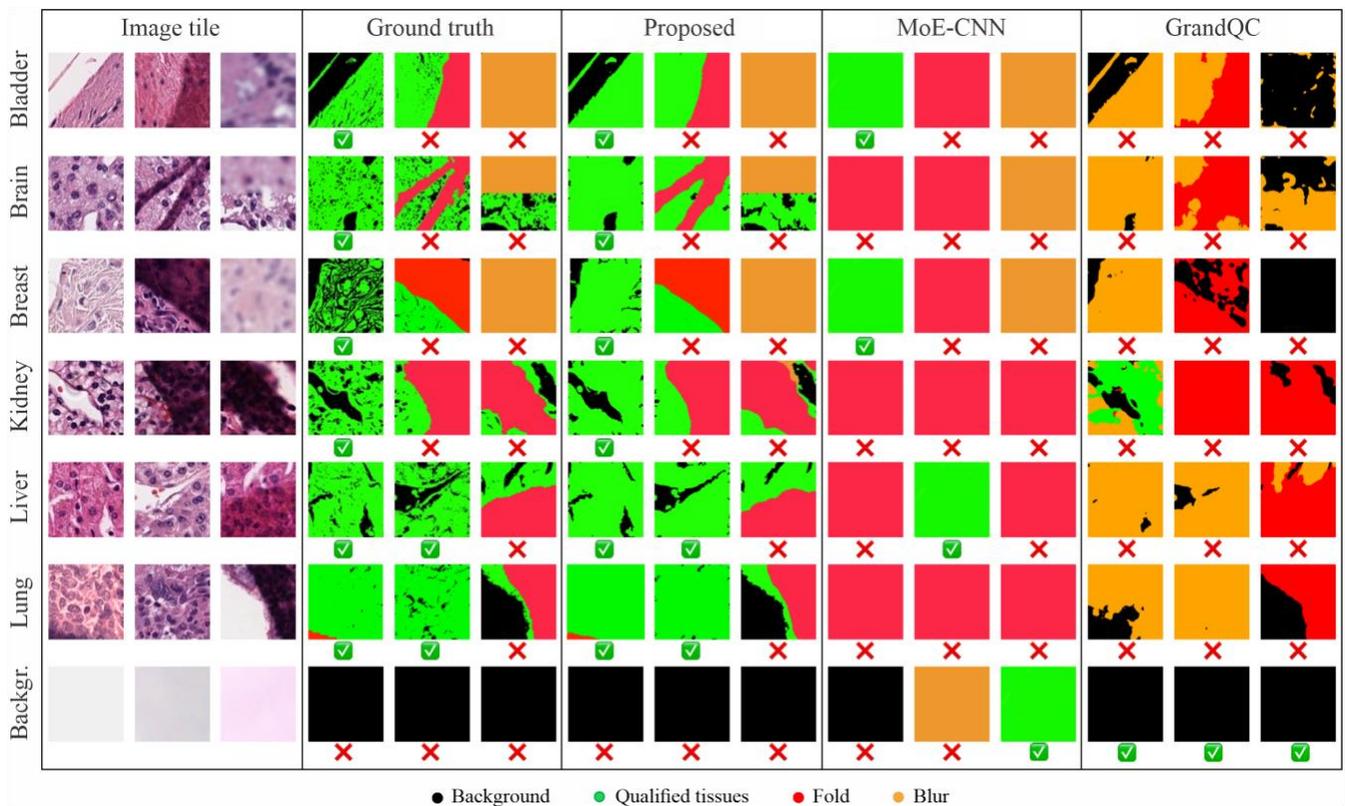

**Fig. 5. Performance visualization:** Comparison of tile segmentation and classification between the proposed model and the state-of-the-art models, MoE-CNN and GrandQC. The image tiles are taken from six different organs (bladder, brain, breast, kidney, liver, lung), as well as background-only tiles. Ground truth annotations are presented alongside model predictions, with tiles labeled as either ✅ (qualified) or ❌ (unqualified). Correct classifications are indicated when the ground truth and prediction labels match. For WSI level see also **Fig.S5** and **Fig.S6**.

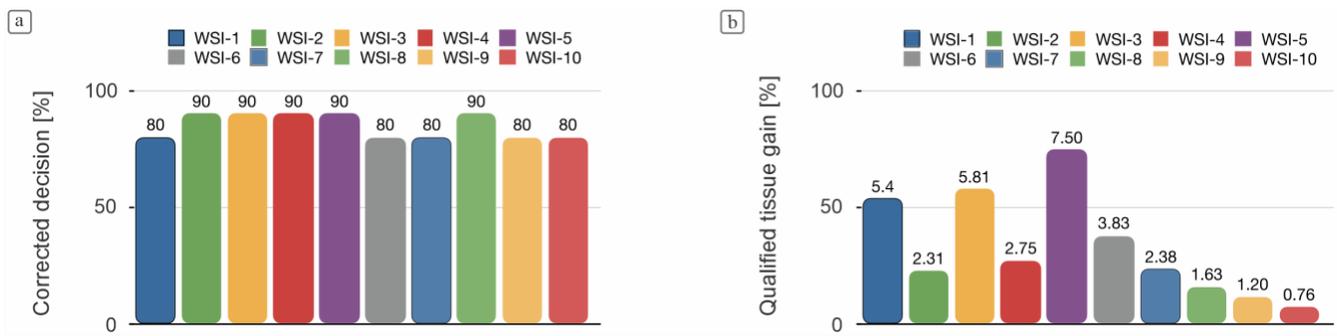

**Fig. 6. Model decision accuracy and tissue gain:** a) Percentage of agreement between the pathologist's selections and the WSI-SmartTiling pipeline's selections; b) Percentage of qualified tissue gain, computed by (3, **method 3.3**), achieved for each WSI.

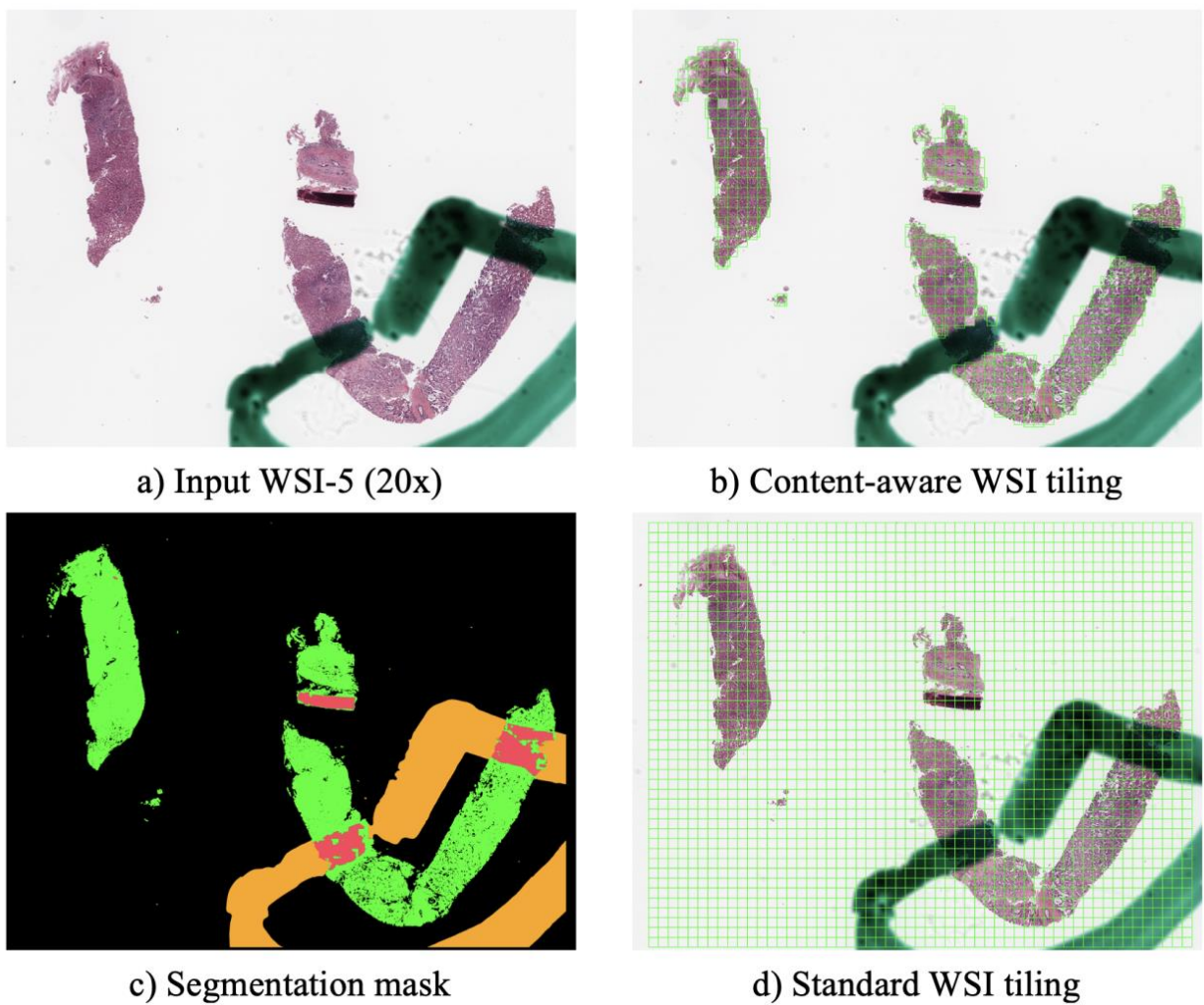

**Fig. 7. Content-aware- vs. conventional tiling:** a) input WSI; b) output of WSI-SmartTiling (Content-aware tiling); c) segmentation mask [background: black, qualified tissue: green, tissue fold: red, blur: orange]; d) Standard WSI tiling (Content-unaware tiling). See also **Fig.S7**. Notably, the artifact segmentation model identifies pen-marking (orange or red) despite not being explicitly trained for pen-marking detection.

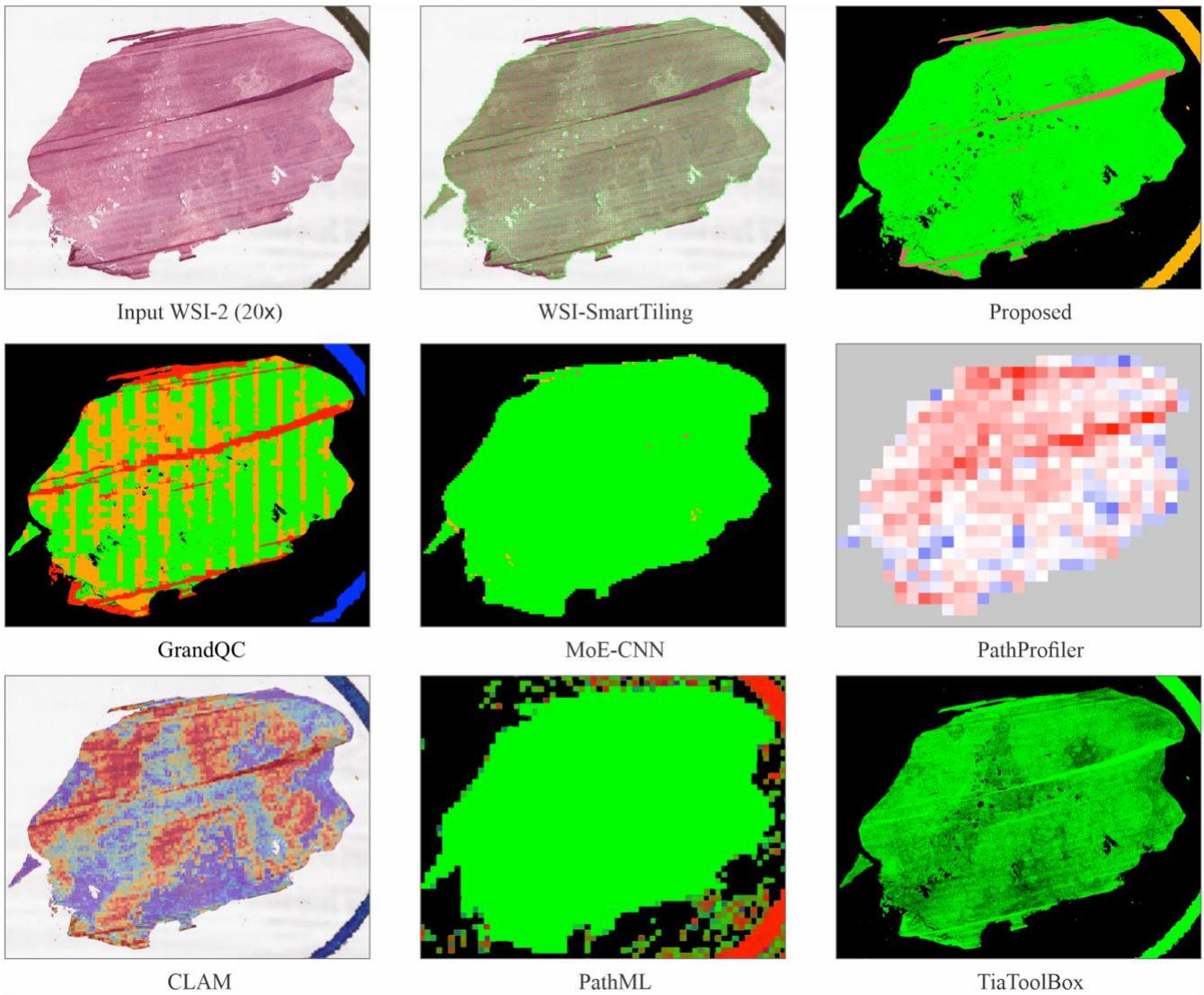

**Fig. 8. Benchmarking visualization:** Comparing WSI-SmartTiling to different pipelines with respect to their support for artifact handling. In GrandQC: red – fold, orange – blur, green – qualified tissue, black – background, blue – pen marking. In MoE-CNN: red – fold, orange – blur, green – qualified tissue, black – background. In PathProfiler: blue – important regions, red – less important regions. In CLAM: red – important regions, blue – less important regions. In PathML: red – pen-marking, green – qualified tissue, black – background. In TiatTooBox: green – qualified tissue, black – background. See also **Fig.S8**.

Table 1. Comparison of features offered by commonly used WSI processing pipelines and the proposed WSI-SmartTiling.

|  | WSI-SmartTiling | GrandQC | MoE-CNN | Pathprofiler | Pathml | HistoROI | CLAM | TiaToolBox |
|---|---|---|---|---|---|---|---|---|
| Tissue Mask | ✓ | ✓ | ✓ | ✓ | ✓ | ✓ | ✓ | ✓ |
| Fold | ✓ | ✓ | ✓ | ✓ | ✗ | ✗ | ✗ | ✗ |
| Blur | ✓ | ✓ | ✓ | ✓ | ✗ | ✗ | ✗ | ✗ |
| Pen-Maker Detection | ✓ | ✓ | ✗ | ✗ | ✓ | ✓ | ✗ | ✗ |
| Pen-Maker Removal | ✓ | ✗ | ✗ | ✗ | ✗ | ✗ | ✗ | ✗ |
| Tiling Optimization | ✓ | ✗ | ✗ | ✗ | ✗ | ✗ | ✗ | ✗ |
| Pixel-wise segmentation | ✓ | ✓ | ✗ | ✗ | ✗ | ✗ | ✗ | ✓ |
| MM. Support (x) | 20, 40 | 5,7,10 | 40 | 5 | 40 | 10 | AD | 40 |
| Targeted Application | preprocess WSI | preprocess WSI | preprocess WSI | preprocess WSI | CP, preprocess WSI | preprocess WSI | CP | preprocess WSI |

Abbreviations – MM: maximum magnification support; DL: deep learning; CP: computational pathology; WSI: whole slide image; AD: indicates as-desired.

# Supplementary Materials

## A. Dataset Creation

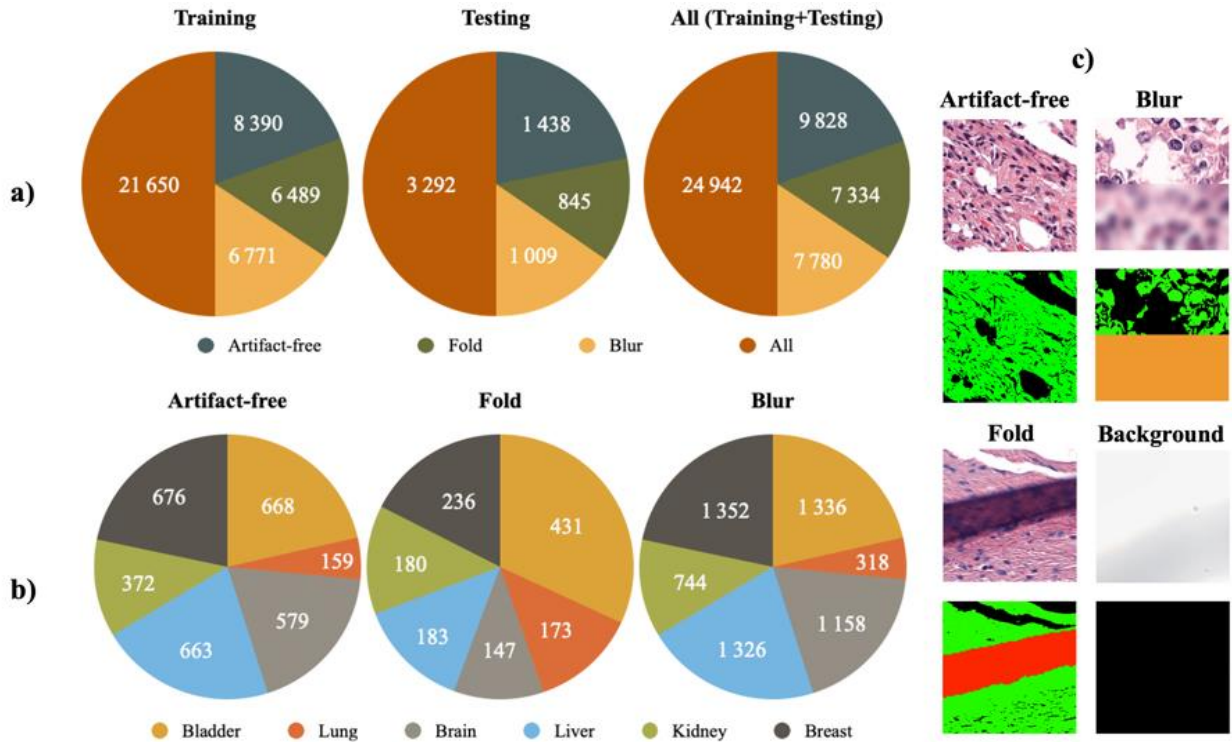

**Fig. S1**. a) Number of tiles for the development dataset: 87% for training, 13% for testing, 100% all (training + testing), and three classes artifact-free, fold, and blur; b) number of tiles for external validation dataset: three classes artifact-free, fold, and blur, across six different organs; c) Example of ground truth tiles with segmentation masks.

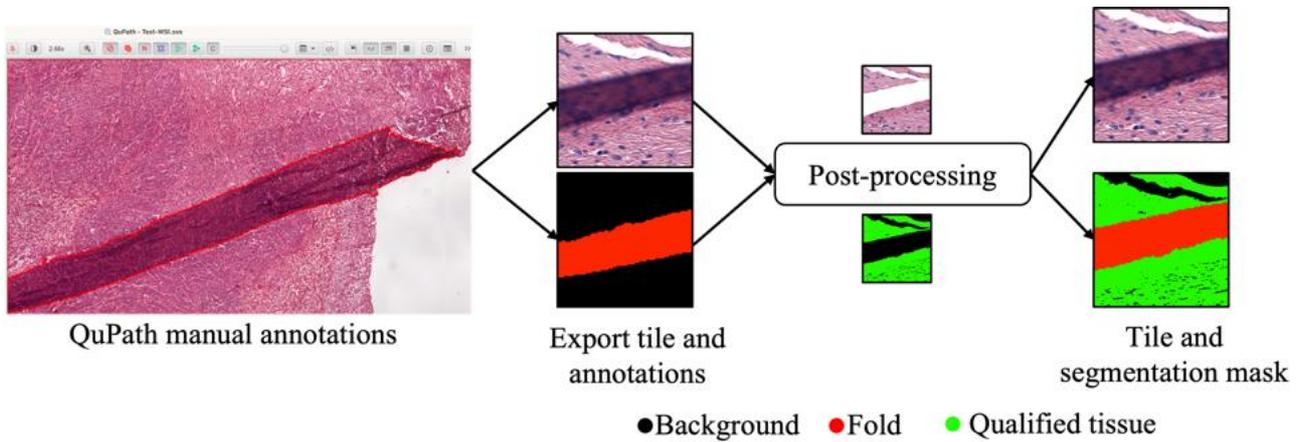

**Fig. S2.** WSI annotation procedure. Artifacts regions are annotated, and then annotated WSIs regions are tiled. Tiles include both annotated (artifacts) and unannotated regions (background and qualified tissues). Unannotated regions are post-processed to separate the qualified tissues from the background. The final segmentation mask has labels: background (0), tissue (1), folds (2), and blur (3).

## B. Artifact Detection Model Architecture

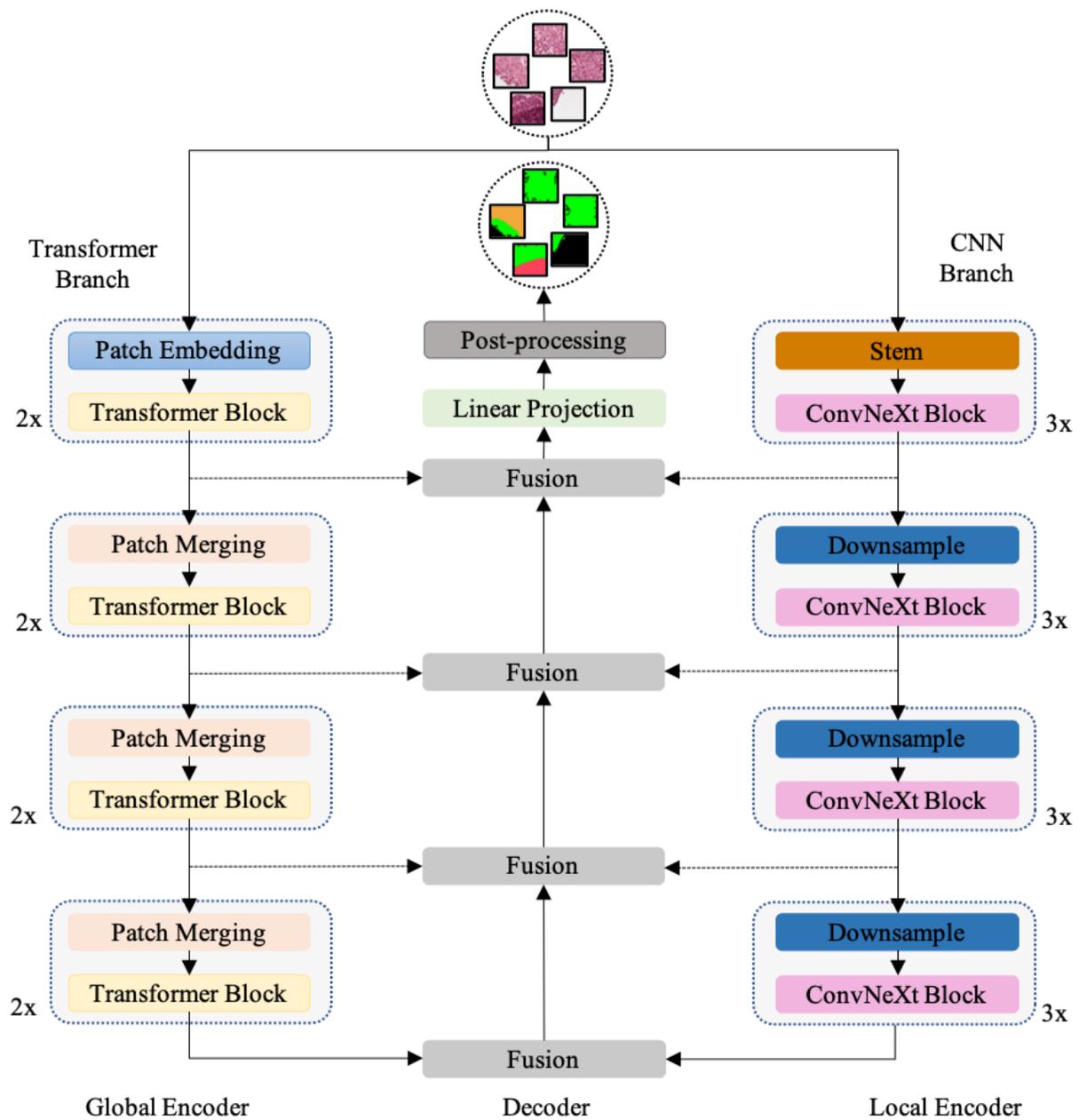

**Fig. S3**. The architecture of the artifact detection model, which is composed of the global encoder, local encoder, decoder, and skip connections.

# C. Additional Results for Artifact Detection Model

**Table.S1.** Confusion matrix for the proposed artifact detection model and benchmark models (Kanwal, et al., 2024) on development dataset test set and the external validation dataset.

**Development Dataset**

Proposed:

| | Artifact-free | Blur | Fold |
|---|---|---|---|
| Artifact-free | 1437 | 0 | 1 |
| Blur | 0 | 1009 | 0 |
| Fold | 2 | 0 | 843 |

MoE-CNN:

| | Artifact-free | Blur | Fold |
|---|---|---|---|
| Artifact-free | 1348 | 0 | 90 |
| Blur | 49 | 747 | 213 |
| Fold | 481 | 10 | 354 |

MoE-ViT:

| | Artifact-free | Blur | Fold |
|---|---|---|---|
| Artifact-free | 173 | 1 | 1264 |
| Blur | 91 | 847 | 71 |
| Fold | 88 | 53 | 704 |

Multiclass-CNN:

| | Artifact-free | Blur | Fold |
|---|---|---|---|
| Artifact-free | 1408 | 4 | 26 |
| Blur | 103 | 904 | 2 |
| Fold | 601 | 47 | 197 |

Multiclass-ViT:

| | Artifact-free | Blur | Fold |
|---|---|---|---|
| Artifact-free | 1331 | 2 | 105 |
| Blur | 731 | 277 | 1 |
| Fold | 447 | 67 | 331 |

**External Validation Dataset**

Proposed:

| | Artifact-free | Blur | Fold |
|---|---|---|---|
| Artifact-free | 3102 | 0 | 15 |
| Blur | 0 | 6234 | 0 |
| Fold | 110 | 8 | 1232 |

MoE-CNN:

| | Artifact-free | Blur | Fold |
|---|---|---|---|
| Artifact-free | 3070 | 0 | 47 |
| Blur | 290 | 3604 | 2340 |
| Fold | 461 | 29 | 860 |

MoE-ViT:

| | Artifact-free | Blur | Fold |
|---|---|---|---|
| Artifact-free | 2680 | 9 | 428 |
| Blur | 1129 | 4377 | 728 |
| Fold | 167 | 47 | 1136 |

Multiclass-CNN:

| | Artifact-free | Blur | Fold |
|---|---|---|---|
| Artifact-free | 3019 | 11 | 87 |
| Blur | 812 | 5412 | 10 |
| Fold | 721 | 83 | 546 |

Multiclass-ViT:

| | Artifact-free | Blur | Fold |
|---|---|---|---|
| Artifact-free | 2970 | 0 | 147 |
| Blur | 2845 | 3374 | 15 |
| Fold | 489 | 26 | 835 |

**Table.S2.** Confusion matrix for the proposed artifact detection model and the state-of-the-art model MoE-CNN (Kanwal, et al., 2024) on external validation set, evaluating results for each organ.

**Proposed**

Bladder:

| | Artifact-free | Blur | Fold |
|---|---|---|---|
| Artifact-free | 668 | 0 | 0 |
| Blur | 0 | 1336 | 0 |
| Fold | 7 | 3 | 421 |

Brain:

| | Artifact-free | Blur | Fold |
|---|---|---|---|
| Artifact-free | 565 | 0 | 14 |
| Blur | 0 | 1158 | 0 |
| Fold | 2 | 3 | 142 |

Breast:

| | Artifact-free | Blur | Fold |
|---|---|---|---|
| Artifact-free | 676 | 0 | 0 |
| Blur | 0 | 1352 | 0 |
| Fold | 66 | 2 | 168 |

Kidney:

| | Artifact-free | Blur | Fold |
|---|---|---|---|
| Artifact-free | 372 | 0 | 0 |
| Blur | 0 | 744 | 0 |
| Fold | 37 | 1 | 142 |

Liver:

| | Artifact-free | Blur | Fold |
|---|---|---|---|
| Artifact-free | 662 | 0 | 1 |
| Blur | 0 | 1326 | 0 |
| Fold | 4 | 4 | 175 |

Lung:

| | Artifact-free | Blur | Fold |
|---|---|---|---|
| Artifact-free | 159 | 0 | 0 |
| Blur | 0 | 318 | 0 |
| Fold | 4 | 2 | 167 |

**MoE-CNN**

Bladder:

| | Artifact-free | Blur | Fold |
|---|---|---|---|
| Artifact-free | 660 | 0 | 8 |
| Blur | 115 | 850 | 371 |
| Fold | 216 | 4 | 211 |

Brain:

| | Artifact-free | Blur | Fold |
|---|---|---|---|
| Artifact-free | 579 | 0 | 0 |
| Blur | 101 | 908 | 149 |
| Fold | 108 | 1 | 38 |

Breast:

| | Artifact-free | Blur | Fold |
|---|---|---|---|
| Artifact-free | 651 | 0 | 25 |
| Blur | 17 | 903 | 432 |
| Fold | 6 | 2 | 228 |

Kidney:

| | Artifact-free | Blur | Fold |
|---|---|---|---|
| Artifact-free | 363 | 0 | 9 |
| Blur | 16 | 401 | 327 |
| Fold | 73 | 2 | 105 |

Liver:

| | Artifact-free | Blur | Fold |
|---|---|---|---|
| Artifact-free | 658 | 0 | 5 |
| Blur | 23 | 428 | 875 |
| Fold | 118 | 1 | 64 |

Lung:

| | Artifact-free | Blur | Fold |
|---|---|---|---|
| Artifact-free | 159 | 0 | 0 |
| Blur | 18 | 114 | 186 |
| Fold | 12 | 1 | 160 |

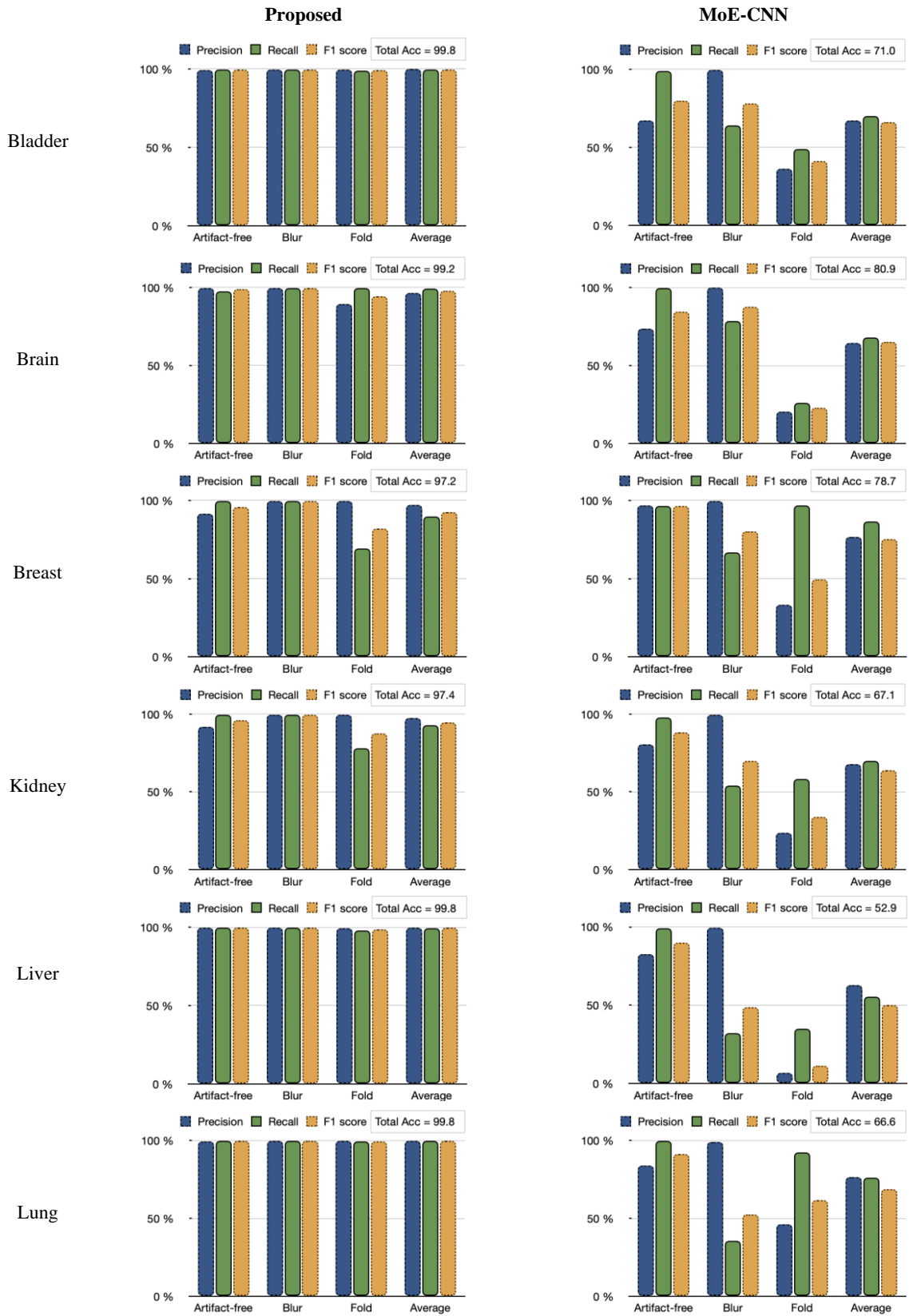

**Fig. S4**. Comparing the performance of the proposed artifact detection model with the state-of-the-art model (Kanwal, et al., 2024) on the external validation dataset, evaluating results for each organ. The last three columns represent the average performance across artifact-free, blur, and fold categories. The proposed model outperforms MoE-CNN across all organ types. It consistently achieves performance for all metrics above 95% across all organs. In contrast, MoE-CNN shows noticeable variation in performance across different organ types, with total accuracy ranging from 52% (liver) to 80.9% (brain). This indicates that the MoE-CNN model cannot generalize well for different tissue types. Furthermore, it has poor performance for fold class. The confusion matrix for each organ and both proposed, and MoE-CNN is provided in Table S2.

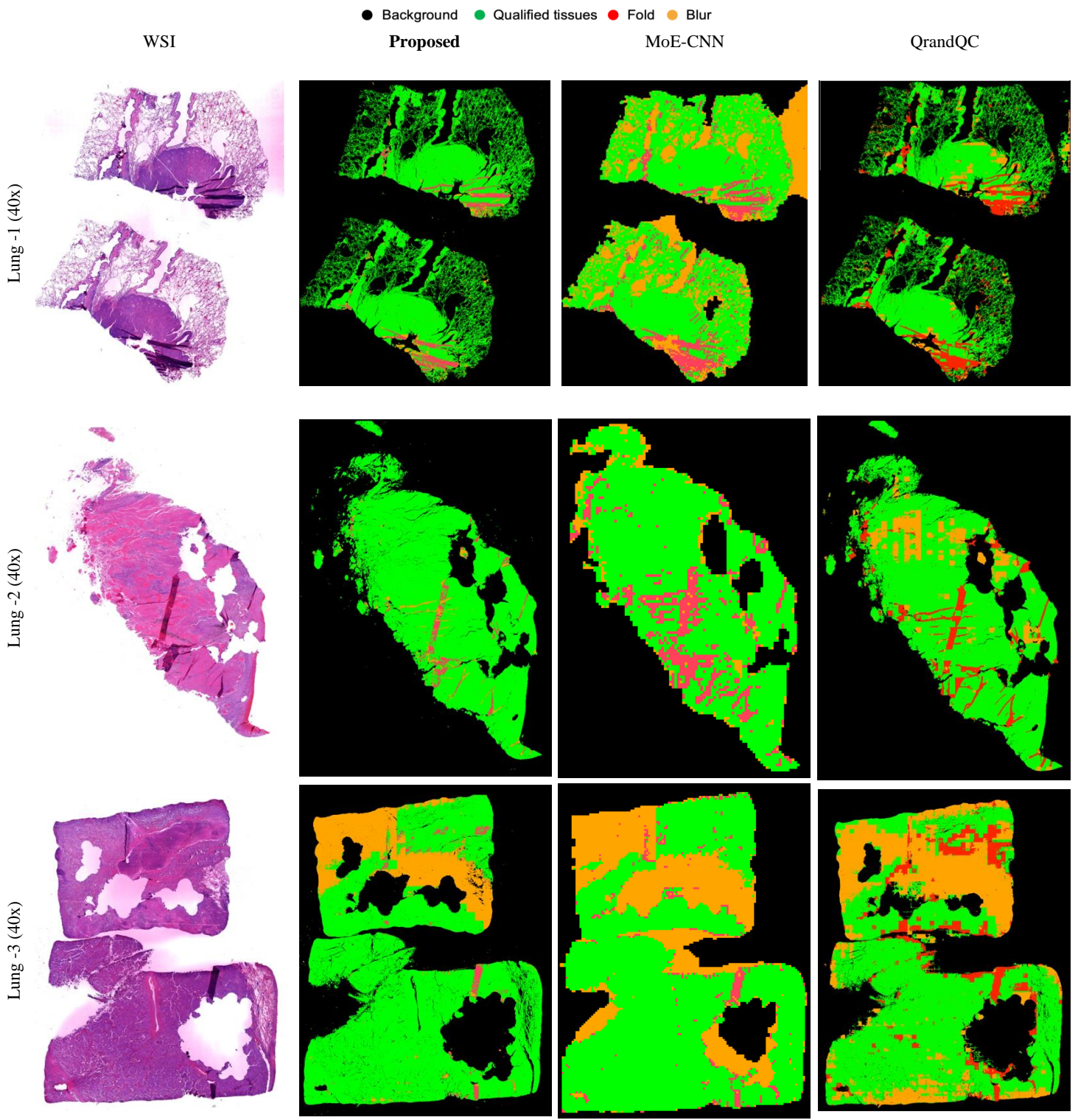

**Fig. S5**. WSI segmentation results for the proposed model compared with the state-of-the-art models MoE-CNN and QrandQC on three WSIs from the development dataset test set. The proposed model demonstrates more precise artifact detection, accurately distinguishing folds, blur, qualified tissue, and background regions, outperforming MoE-CNN and QrandQC. Note that QrandQC performs very poorly at 40x magnification (mostly blurred regions); therefore, all results were obtained at 20x magnification. The WSIs and segmentation masks with high resolution are available in SupplementaryMaterial.

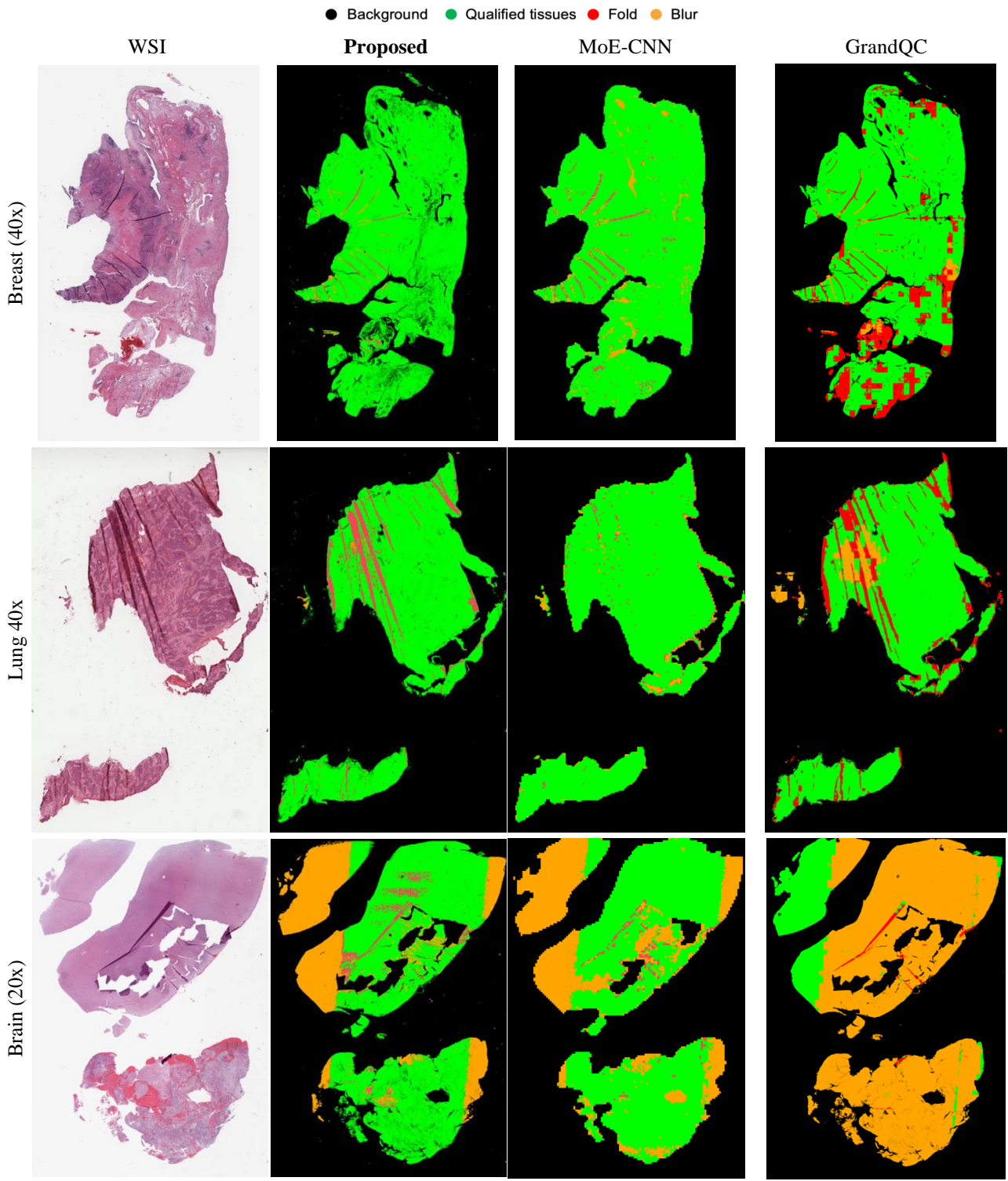

**Fig. S6**. WSI segmentation results for the proposed model and the state-of-the-art models MoE-CNN and QrandQC on Breast, Lung, and Brain WSIs from the external validation dataset. The proposed model accurately detects folds, blur, background, and qualified tissue regions, while MoE-CNN struggles with misclassifications, particularly regions with folds, background, and qualified tissue. Similarly, QrandQC often misclassifies qualified tissue as folds or blur, especially at 20x magnification. The WSIs and segmentation masks with high resolution (including Bladder Kidney Liver organs) are available in SupplementaryMaterial.

# D. Additional Results for WSI-SmartTiling Pipeline

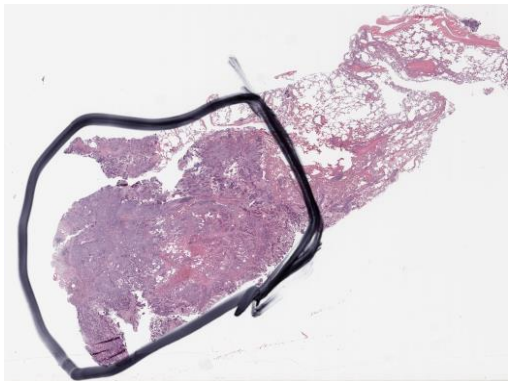

Input WSI-1(20x)

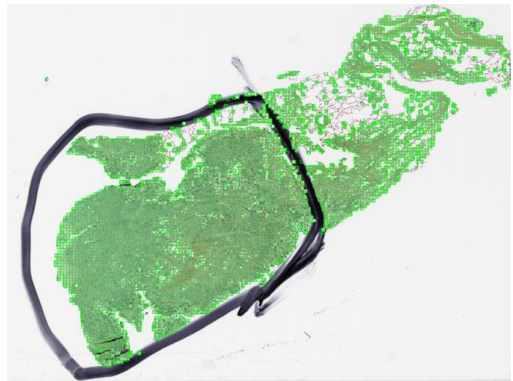

WSI-SmartTiling: Content-aware tiling

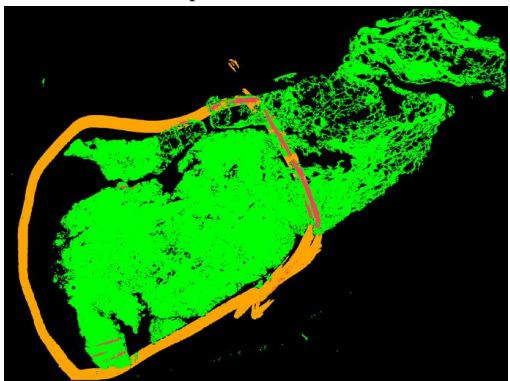

Segmentation mask

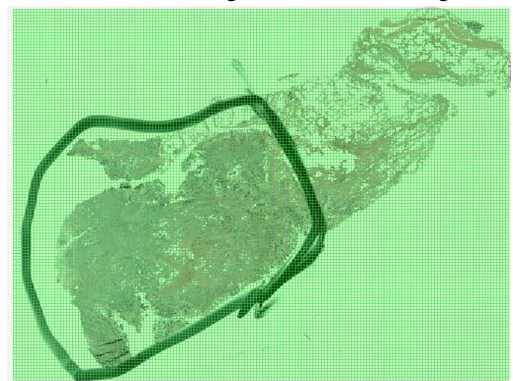

Standard tiling

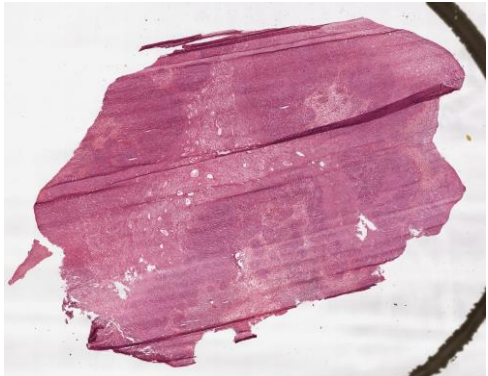

Input WSI-2 (20x)

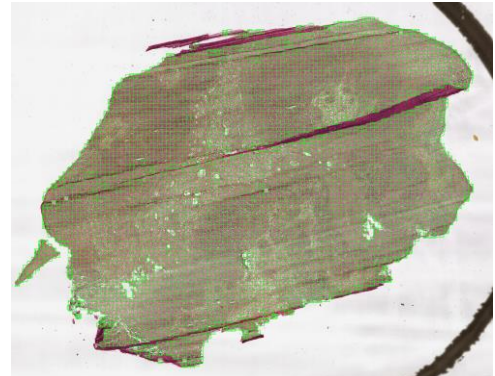

WSI-SmartTiling: Content-aware tiling

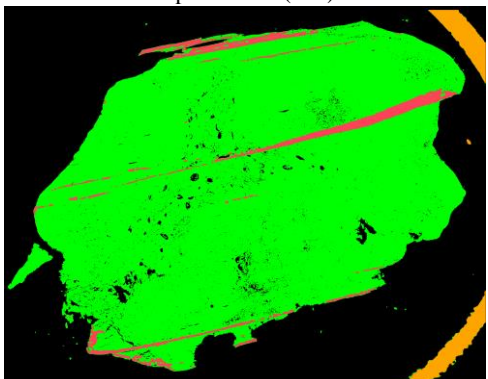

Segmentation mask

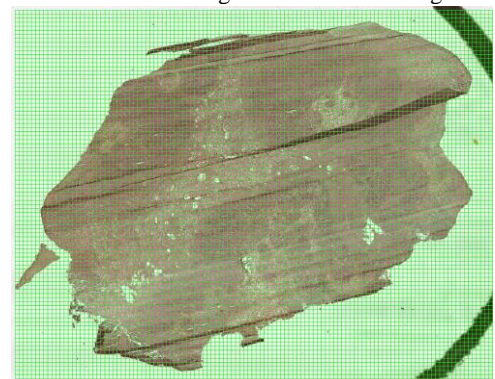

Standard tiling

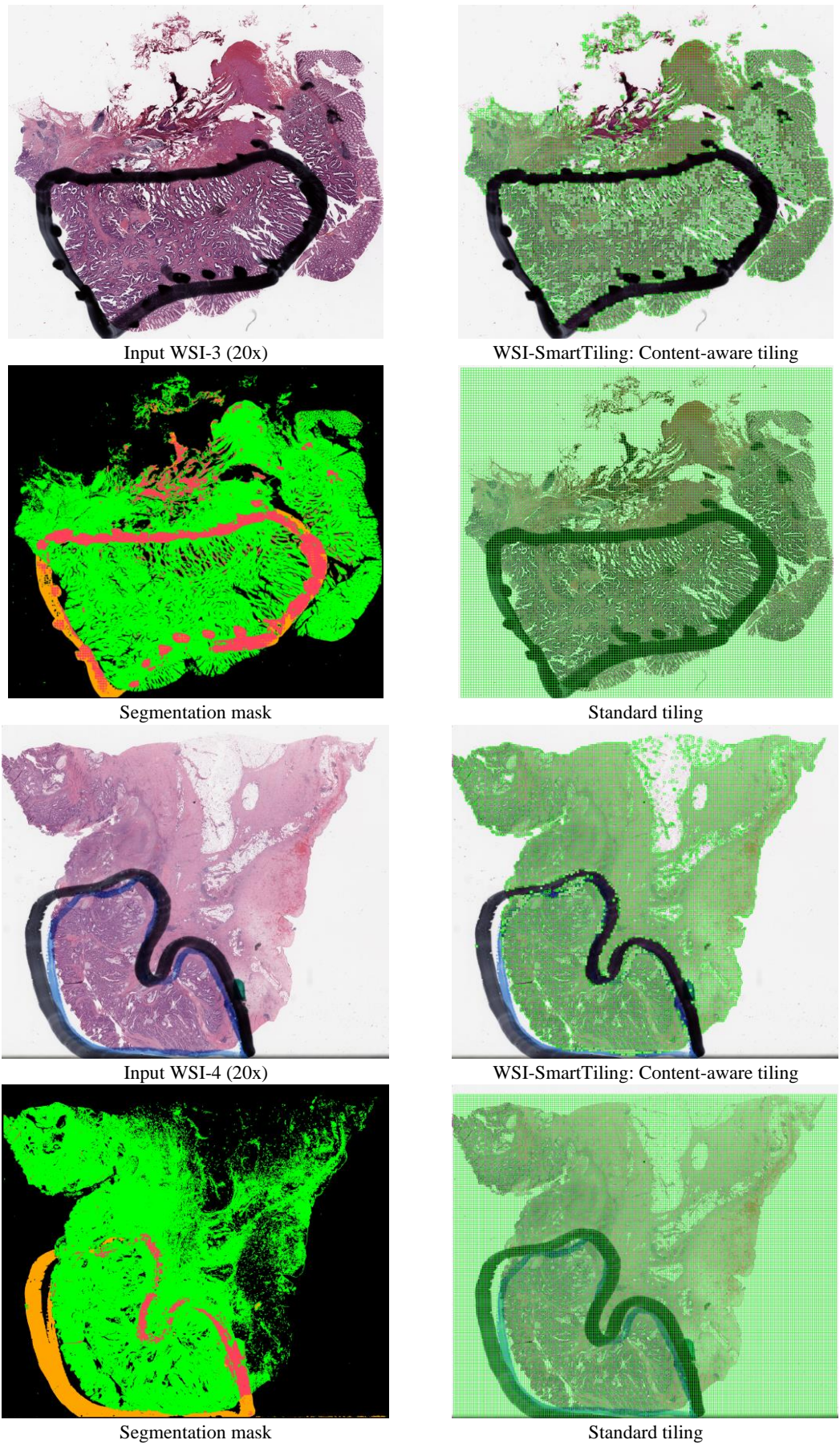

**Fig. S7.** Comparison of content-aware tiling and standard (content-unaware) tiling. The proposed WSI-SmartTiling effectively selects qualified tissue regions, excludes artifacts, and minimizes the exclusion of diagnostically important tissue. In the segmentation maps: background (black), qualified tissue (green), folding (red), and blurring (orange). Notably, the artifact segmentation model identifies pen-marking (orange or red) despite not being explicitly trained for pen-marking detection. The WSIs with high resolution are available in SupplementaryMaterial.

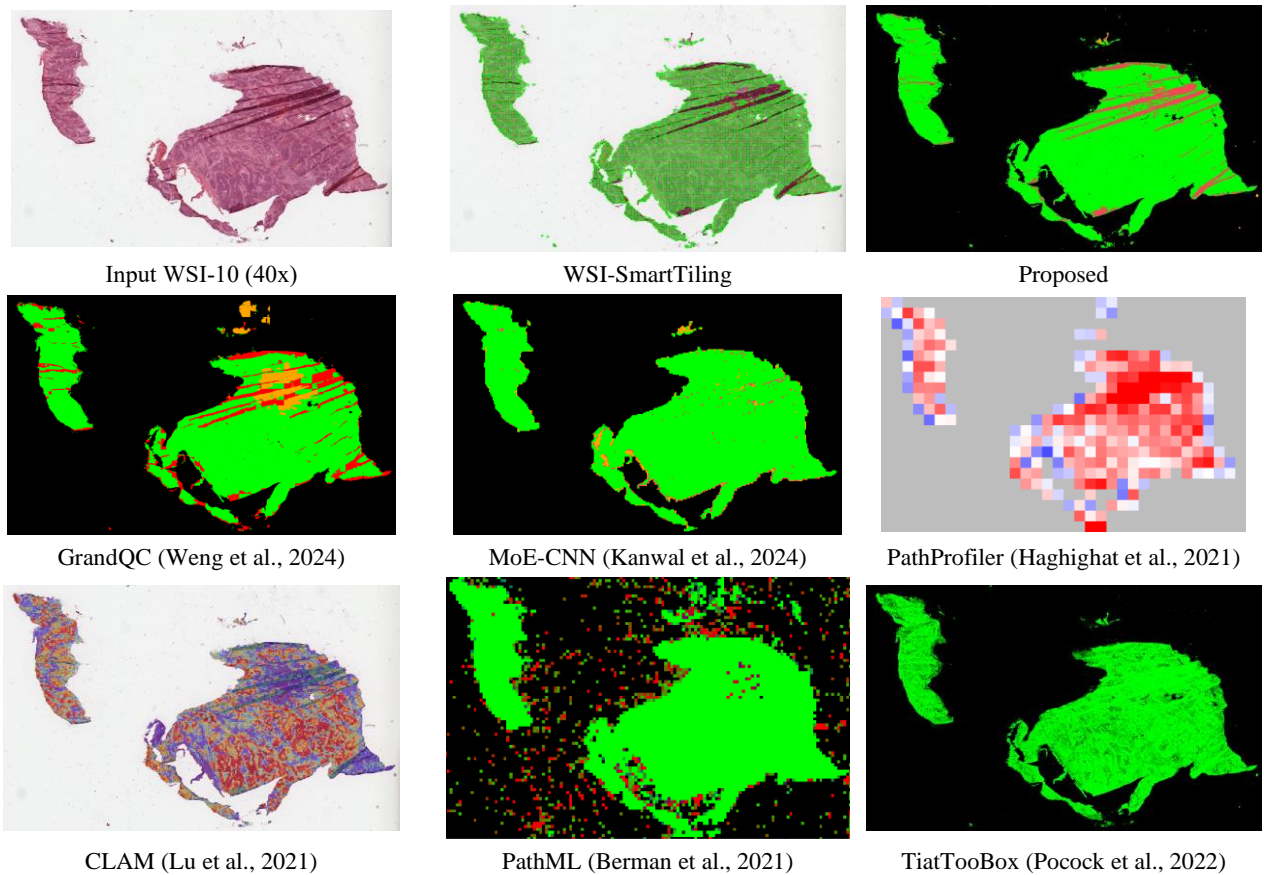

**Fig. S8.** Comparing WSI-SmartTiling to different pipelines with respect to their support for artifact handling. GrandQC and MoE are specifically designed to detect artifacts in WSIs. However, as can be seen, GrandQC struggles to accurately identify folding artifacts, often misclassifying qualified tissue as folding or blurring artifacts, and MoE struggles to detect folding artifact. PathProfiler and CLAM generate scoring maps that highlight important WSI regions in red (blue for PathProfiler) and less important regions in blue (red for PathProfiler). However, they lack the capability for artifact detection. PathML generates a segmentation mask separating tissue from background, but it cannot detect other artifacts. TiatToolBox also generates a tissue-background segmentation mask (better than PathML) but offers no artifact detection capabilities. HistoROI does not support WSIs with magnifications greater than 10x.